# EDUCATION AND SUSTAINABILITY: A MODEL FOR DIFFERENT ENGINEERING DEGREES

**J. Bilbao    E. Bravo    O. Garcia    C. Rebollar**

*Applied Mathematics Department, University of the Basque Country (UPV/EHU), Bilbao, Spain*
*javier.bilbao@ehu.eus, eugenio.bravo@ehu.eus, olatz.garcia@ehu.eus, carolina.rebollar@ehu.eus*

**Abstract-** Technologies related to the Internet of Things (IoT) have seen remarkable growth in recent years. This has facilitated, among many other reasons, that monitoring systems have spread in many everyday areas, including both industry and the services and systems of the so-called smart home. These systems can also be applied in Engineering and in Education for the different existing engineering degrees; and one of the fields is sustainability. A project related to sustainability and student practices has been launched at our university. In this way, several objectives are achieved at the same time, such as the transfer of knowledge from universities to society, and also the development of sustainable education, in line with the sustainable development goals. In this framework, we want to apply the ideas of monitoring through IoT applications, by means of the measurement of certain environmental factors that occur both in an urban garden and in a composting process. Only open hardware-based devices have been used in the project. The proposed model can be applied in other areas of knowledge, having considered different alternatives and having chosen the best elements, based on sustainability criteria, for each section of the project. Specifically, in the system that has been created, the environmental factors of a small urban garden and also of a composting box can be measured. Both sections of the project, garden and composting, are located at the university. The factors to be measured are the following: air temperature, air humidity, soil moisture, ultraviolet radiation and amount of light (luminosity) received in the urban garden; and temperature and humidity in the composting process.

**Keywords:** IoT Technologies, Education, Engineering, Sustainability, Urban Garden.

## 1. INTRODUCTION

When we speak of Education and Sustainability, some educational models offer guidelines and normative schemes for educational intervention, articulated in well-founded methodological proposals, and in sequences aimed at clarifying the idea that the actions of human societies on the area of study are of a normative nature [1]. In other words, it is assumed that these actions are mediated by interests that are legitimized in ethical, political or economic discourses, in step with the evolution of the conditions of human production and reproduction. From a theoretical point of view, Gutiérrez distinguishes two modalities of curricular integration in the field of formal education: one of an interdisciplinary nature, and the other of a multidisciplinary nature [2]. In our case, the development of our activities is framed within the multidisciplinary modality, having a cross-curricular character and being able to be applied in different engineering degrees.

The Internet of Things (IoT) is being introduced in all areas of society, and one part of it is the measurement of environmental factors. Although this observation and measurement task is an action that human beings have always carried out, both in ancient times and nowadays. Knowing the values of different factors helps to obtain better yields in different fields. For example, in agriculture, it is important to know the amount of light a plant receives or the degree of soil moisture in order to irrigate at the optimum time, so that water is not wasted and there is no risk of the plant drowning. Composting is a well-known technique that contributes to reducing the pollution generated by waste and, therefore, to preserving the environment. In addition to recycling organic waste, natural compost can be produced both for personal gardens and to nourish the soil of agricultural land. But the process is very sensitive to temperature and humidity.

In addition, on the emergence of IoT (Internet of Things) systems, there are many companies and individual users who use these factors to measure, install monitoring systems in different locations and collect data. This has several advantages, since the remote control and visualization of the data allows to know the nature of the elements to be controlled anywhere and at any time. This project will therefore be carried out taking into account the relevant aspects in terms of sustainability. For this, it is intended to measure five environmental factors (temperature, air humidity, ground humidity, ultraviolet radiation and luminosity) of a small orchard located at the university, measure the temperature and humidity factors that occur in a compost box that will be next to it, collect them through free hardware components, store them and visualize them through an internet platform.

In addition, to do all this is intended, deepening in the field of sustainability, to feed a free hardware board that will be used and its sensors through an autonomous power supply system. If we take advantage of the benefits offered by renewable energies and free hardware, it is intended to access the world of IoT systems and control the activity of environmental conditions and composting of an orchard through the monitoring system to be developed. Therefore, the results of this project and the generated system can be further developed in future projects.

## 2. SUSTAINABILITY EDUCATION FIELD

Sustainability education currently promoted and developed does not have a unique manifestation, nor does it conform to an exclusive prototype of characteristic educational intervention; rather, there are diverse practices guided by divergent interests, mediated by resources, contexts and instruments of varied nature, and promoted by agents of heterogeneous character. Thus, according to Sauvé et al. sustainability education is part of a long historical trajectory, through which it has acquired a triple relevance: social, environmental and educational.

According to the same authors, various currents of thought and practices have emerged, determined by the ideological and ethical roots of the various protagonists, and by the different representations of education, the environment and development that they adopt [3].

Along the same theoretical line, other authors share the idea of seeing sustainability and the way of bringing it to the educational system as an amalgam of initiatives with different degrees of intentionality and with a plurality of execution and implementation modalities. This progress is possible thanks to the theoretical effort we have been making in recent years on the subject from different communities of practice [4-10].

The confluence and mutual enrichment of anthropological, psychological, industrial, sociological, economic, ecological, etc., knowledge suggest that it is feasible to construct a coherent and complex image of the contemporary ecological crisis [1]. Not only can it be formalized, but it can be interpreted and rationalized by human thought. This leads to a normative orientation to be adopted by the changes that will make it possible to overcome it, and also to define the meaning and role of the educational system as part of an education with the vocation of moving towards the integration of human development in the coordinates of a progressive reconciliation with the environment [11, 12].

According to the consulted references, we can say that there are five types of theoretical contributions that can contribute to organize the multiple models in force: economic, socio-political, philosophical and bioethical, pro-environmental, and pedagogical. Each type emphasizes the disciplinary field and scientific culture from which it emerges. In addition, their arguments are based on the concepts, traditions, theories and languages of the discipline in which the author (or authors) making the contribution has been trained.

## 3. DATA ACQUISITION SYSTEMS

In the same way that humans can sense environmental conditions through the senses, electronic systems can receive this environmental information through DAQ (Data Acquisition) systems. Also, information from these electronic systems can be used by humans for other purposes.

Nowadays, they have become a fundamental part of human life and we can observe them in any type of activity, such as the control of industrial processes, cars, airplanes, medical equipment, household appliances, etc. In the field of agriculture, the situation is not much more unequal, since the use of sensors to control different factors and applications is increasing. Among these applications, for example, would be automatic irrigation, which is based on the right time to irrigate by controlling soil moisture. Other applications would be: control of the internal conditions of the greenhouses, control of luminosity, temperature, ventilation, humidity, etc., control of the level of ultraviolet radiation, etc. This would lead some studies to consider sensor networks as one of the 10 technologies that would change the world for the better [13].

In order to understand the information received through the sensors, a processor is needed to translate the electrical or physical phenomena received from the sensors (change of voltage, current, temperature, pressure) into understandable data. Therefore, DAQ systems are used to carry out this process. These DAQ systems consist of sensors, a data acquisition hardware and a PC with programmable software. Compared to traditional measurement systems, these new PC-based systems offer greater processing, productivity, visualization and connectivity possibilities, thus making measurement systems more efficient, powerful and cost-effective [14].

If each of the sections of these systems is analyzed, the following can be said:
- Sensors

Sensors, also known as transducers, are responsible for reading a physical quantity, such as temperature, and converting this reading into an electrical signal that can then be measurable and communicable.
- DAQ devices

This device is the one that allows the interaction between the computer and the user's sensors. Its main function is to digitize the data it receives from the sensors so that it can be understood by the computer. To carry out this function, they have a circuit to adapt the signals received to a communication protocol, a converter to be able to pass the signals from analog to digital and finally a bus to communicate with the computer. This communication can be of several types: Ethernet, Wi-Fi, ModBus or others. There are DAQ devices that have a viewer to view the collected data, but often they do not, and, in these cases, we can only see it when the information reaches the computer.

Among this type of devices, we can find those known as process loggers or dataloggers, which are usually based on microcontrollers. These systems usually have internal memory, which gives them autonomy for data collection, allowing the data received from the sensors to be stored there. Therefore, they are very useful as a data collection system. This function is very useful, since if at any time the communication with the computer was interrupted, the system would have been able to store the data in the internal memory and not lose information. Therefore, the biggest advantage that these dataloggers have would be their ability to function even without the need for a computer. Because of this advantage, in many projects these types of DAQ systems are used as data collection systems.

Data loggers can be of various types. Some are capable of interpreting only one sensor, being very cheap, but others can have hundreds of inputs and have a great power in both software and hardware. For this project we will use a data collection system of this type, as it allows us to cover many different applications for the agricultural sector.

- PC with Inspection Software

Finally, to carry out the whole system, there would be a PC with all the software possibilities. This computer allows the interaction between the end user and the system. In addition, it enables to visualize, collect the information received, process it freely and perform the necessary calculations. The computer can also be used to control the configuration and operation of the entire system.

On the other hand, it should be noted that the use of these systems has greatly increased in areas such as industry and mining, which offer many advantages. For example, they facilitate data analysis and help make information more accessible, especially when dealing with large volumes of data. In addition, they offer a better level of data security, since as soon as data is received it can be sent and saved to different devices so that it is not lost. On the other hand, they offer great facilities and support for monitoring processes, allowing sensor networks to be physically located in different places and having the capacity to cover large areas. In this way, these systems are capable of capturing and monitoring at all times data from an infinite number of different parameters in large areas. Finally, since each of the above components can be modified or upgraded independently, cost savings are achieved.

## 3. CAMPUS BIZIA LAB AND SERVICE-LEARNING

The Campus Bizia Lab (CBL) program is an initiative derived from the Erasmus University Educators for Sustainable Development Project in which the University of the Basque Country (UPV/EHU) participated between 2013-2016. It aims to trigger a collaborative process between faculty, administration and services staff and students (transdisciplinary approach) in order to respond to sustainability challenges within the University itself.

Campus Bizia Lab consists of a research/action process that aims to develop a high impact practice among students (transdisciplinary learning based on sustainability challenges) in which faculty act as researchers of their own practice. This high impact practice has a curricular nature and is materialized through the Final Degree Project (FDP) and Final Master's Project (FMP).

The main objective of the program is to respond to the SDGs proposed by UNESCO within the university itself, through the FDPs and FMPs of different degrees. At the same time, it seeks to articulate transdisciplinary communities where different figures working specifically in the field of education for sustainability are involved in a cooperative manner: students, faculty and administrative and service staff. In this way, the different projects that make up the CBL create synergies focused on the resolution of challenges and problems of unsustainability that are detected in the university campuses themselves, generating multilevel sustainable practices. The transdisciplinary nature of the program means that the learning processes linked to the design, development and evaluation of the projects have a high curricular impact on the students.

Service-Learning (SL) is a methodology based on practical and reflective experiences in which students are involved in a context with real problems, thus developing content and professional skills [15, 16]. One of the characteristics of SL is that the real needs of the community are detected, or at least some of them, normally related to the academic area being taught, in such a way that the link between educational institutions is promoted. Thus, university students can improve their training by connecting directly with the social reality [17]. Another advantage mentioned about service-learning is that it also creates and strengthens bridges of coordination between the university and the educational community [18-22]. Typically, projects are based on the following concepts:
➢ Experience
➢ Active participation
➢ Interdisciplinarity
➢ Teamwork

Among the possible actions that can be carried out at the university, one of the most complete is the development of a Final Degree Project (FDP). The FDP is a project, report or study, which is carried out at the end of undergraduate studies. In Spain, in all undergraduate degrees, and after the university reform that was implemented at the end of the first decade of the 21st century (Bologna process), the development and presentation of an FDP is mandatory for all students.

At the UPV/EHU, this methodology fits within its IKD framework of cooperative and dynamic learning (the university's own methodology) and within the application of the SDGs through initiatives such as the aforementioned Campus Bizia Lab.

## 4. OPEN HARDWARE

Open hardware is born from the ideology of open-source software. The latter appeared [23] as early as 1980. The software at that time was mostly proprietary and in order to modify it, several programmers began to encourage the trend to create free software programs.

Thus, over the years, Operating Systems based on free software were also created, such as GNU, and it has been an idea that has been evolving up to the present day.

Free software defends four fundamental ideas: freedom of use, freedom of learning and change, freedom of distribution and dissemination of improvements. However, sharing hardware designs is not as easy as sharing software, since the hardware has to be physically the same to work exactly the same, and when copying the design, manufacturing costs have to be taken into account, and finally, it is necessary to make sure that the materials for the product we want to develop are available.

However, overcoming these problems, the idea of open hardware is very important to create a competitive market. Many people have good ideas, but the materials needed to implement them can be very expensive or they lack the information about the products they want to use to understand how they work. Open hardware corrects these problems relatively well, bringing product prices down and making designs available to anyone. In addition, products based on this idea, when the product is known to be faulty, offer the possibility of fixing it by themselves without having to take it to the store and pay a fortune. In the world of open hardware, you can find many different products, but the best known are Raspberry Pi and Arduino. In this project we will use the latter to create the monitoring system, so we will focus on it.

Arduino is one of the pioneers and most popular open hardware products [24]. It consists of an open hardware board and a microcontroller that provides an environment for developments. It has several analog and digital inputs, where sensors or switches can be connected, and several outputs, to connect motors or switches. All the ingredients are cheap and the user himself can assemble an Arduino board, buy the ingredients and see the plans; but you can also buy the board itself assembled and use this option to develop the project. It is, in short, a small and cheap computer, but in which you can replace the usual inputs (keyboard, mouse) or outputs of the computer (screen, printer) by sensors or motors that the user wants.

Thus, by connecting the necessary sensors and devices to be measured, and writing a program through the development environment, a complete monitoring system can be created, as will be done in this project. Moreover, since you can also find useful modules for the creation of stand-alone projects, it will be highly recommended for the development of the stand-alone product you intend to create.

On the other hand, on the Arduino website you can find numerous helps shared by the community, as well as a lot of developed codes and examples. For all these reasons, we have opted for the use of open hardware in this project, taking advantage of the benefits offered by the Arduino boards and with the idea of developing the project further in the future, leaving both the design and the codes in the hands of anyone:

## 5. METHOD

There are several methodologies used in sustainability education studies. For example, according to Johnson and Manoli [25], the concept of environmental perceptions includes attitudes, concerns, beliefs, paradigms, values and points of view regarding the environment. On the other hand, to study the perception of adults regarding the environment, a methodology called New Ecological Paradigm (NEP) is commonly used [26]. This methodology was modified and validated for use with children and adolescents [27]. However, such methodologies do not determine environmental knowledge or pro-environmental behavior [28, 29] and must take into account more factors, such as the age of the subjects.

Therefore, there is no standard methodology to assess environmental knowledge and pro-environmental behavior. The present study was based on the methodologies proposed by Evans et al. [28] and Kaiser et al. [29]. Both methodologies had to be modified according to the defined objectives. In addition, it was necessary to develop an instrument to evaluate these variables in the context of sustainability education, so a quantitative method based on surveys was used.

## 6. DATA ANALYSIS AND RESULTS

Within the population of 43 students who were going to do the Final Degree Project, the sample was formed by 41 of them. The sample size is adequate with parameters of 50% heterogeneity, a 5% margin of error, and 95% confidence. In addition, we divided this number of students into two groups: a control group, consisting of 24 students; and an experimental group, consisting of 17 students. In the experimental group, the Campus Bizia Lab was presented as a service-learning, focusing part of the teaching on sustainability through urban gardens, composting and all the systems surrounding these actions; while the control group followed a more traditional training. Some of the students in the experimental group carried out their Final Degree Project on the subject of urban gardens, and some of the technical results are presented in the following section.

Two tests were given to all the students: one before all the sessions (pre-test), that is, at the beginning of the course; and another one at the end of the classes (post-test). The pre-test and post-test means are presented in Table 1. All the data refer to values obtained by the experimental group and by the control group, presenting the descriptive statistics achieved by each of them and the total sample. It can be observed that, in the pre-test, the mean for the first group was 11.30 and standard deviation equal to 7.1; for the control group 11.9 and 8.0, respectively. The difference between these means was evaluated through a one-way analysis of variance, and the results of which are shown in Table 2. Values of $F = 0.051$, $p > 0.5$, indicate similar values between the two groups (experimental and control)., in terms of the information requested in the pre-test.

Table 1. Pre-test and post-test descriptive statistics for the experimental and control groups, and for the total sample

| Statistic | Experimental group | | Control group | | Sample | |
|---|---|---|---|---|---|---|
| | Pre-test | Post-test | Pre-test | Post-test | Pre-test | Post-test |
| Participants | 17 | 17 | 24 | 24 | 41 | 41 |
| Mean | 11.3 | 18.7 | 11.9 | 11.8 | 11.6 | 14.6 |
| Standard deviation | 7.1 | 5.7 | 8.0 | 7.9 | 7.55 | 7.79 |
| Coefficient of variation (%) | 62.37.78 | 30.60 | 67.78 | 67.01 | 64.98 | 53.23 |
| Median | 10.3 | 18.5 | 9.8 | 11.8 | 10.2 | 14.5 |

Table 2. Analysis of variance of the pre-test scores for the experimental and control groups

| Source of variation | Sum of squares | Degrees of freedom | Mean square | F | p |
|---|---|---|---|---|---|
| Group | 3.02 | 1 | 3.02 | 0.051 | 0.821 |
| Error | 2280.78 | 39 | 58.48 | | |
| Total | 2283.81 | 40 | | | |

Table 3. Analysis of variance of the post-test scores for the experimental and control groups

| Source of variation | Sum of squares | Degrees of freedom | Mean square | F | p |
|---|---|---|---|---|---|
| Group | 475.01 | 1 | 475.01 | 9.465 | 0.003 |
| Error | 1957.08 | 39 | 50.18 | | |
| Total | 2432.09 | 40 | | | |

As for the means obtained in the post-test (Table 1), we can highlight that for the experimental group it was 18.7, with a standard deviation of 5.7, and for the control group it was 11.8 and 7.9, respectively. This difference between means was also evaluated through a one-way analysis of variance. These results are shown in Table 3. If we observe the values of $F = 9.465$, $p < 0.01$, we can conclude that there is a highly significant difference in favor of the experimental group. According to the results described, the experimental group obtained greater gains and homogeneity in the post-test scores, with respect to the control group, once the implementation of the introduction of sustainability education was developed.

## 7. TECHNICAL RESULTS

The results obtained after the complete configuration and development of the monitoring system will be shown below. First, a simple communication between both stations, configured by means of NRF24L01 modules, will be carried out to guarantee the connection and the correct communication. Then, with the complete code developed, the data will be sent from the remote station to the base station and from there some of the data collected in the Ubidots platform will be shown through the WiFi configuration.

It must be said that this project will not analyze in depth the values of the results obtained, i.e., it will not analyze how the values obtained affect the plantation, nor will it measure the conditions that occur in the composting box throughout the process.

In addition, it would be necessary to implement liners and protections to be able to place them outdoors during adverse weather conditions. Some of the results provided by the system can be seen in the Figures 1 and 2.

• Light level intensity (Lux):

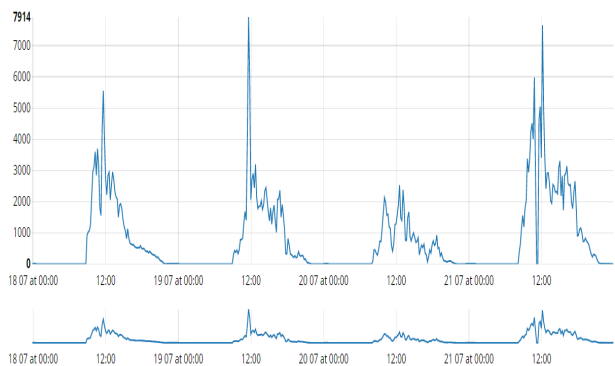

Figure 1. Light level intensity in the urban garden during some days

• Air humidity (%):

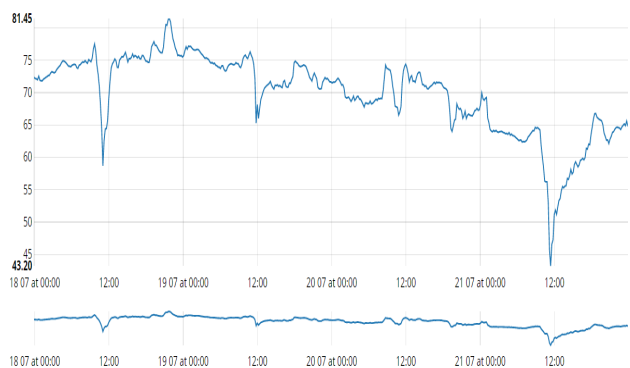

Figure 2. Air humidity in the urban garden during some days

Looking at the graphs, it can be said that the values are well collected and make sense, they are also shown in an easy-to-understand way and the data of the days of the different sections can be seen.

## 8. CONCLUSIONS

The main objective of this project is to develop a monitoring system to measure some environmental factors and the ins and outs of a composting process, using open hardware for this purpose and electrically powered by an autonomous system. To achieve this goal, hardware and software analysis has been carried out, with different options. A complete design of the autonomous power system has also been carried out and an efficient way of wireless communication has been designed. Once the methods and components to be used were selected, we proceeded to the system assembly and code creation section.

To fulfill all these functions, it has been proposed to develop a topology with 2 boards, in which one would work as a base station collecting the data by wireless connection and sending them to the IoT platform via WiFi, and also saving them locally on the SD card. The other would work as a remote station, connecting to it both the orchard and the compost monitoring sensors, collecting data from them and sending them to the base station through a module that allows radio communication. By having to face real situations and to respond and learn to adapt to different educational situations, the university students have also acquired engineering skills.

## ACKNOWLEDGEMENTS

The authors are grateful to Jon Bilbao for his collaboration in the development and implementation of the project. The authors also thank the Sustainability Directorate of the University of the Basque Country (UPV/EHU) for its support and the grant that have made possible this research.

The authors thank the support of the laboratory of the Applied Mathematics Department of the University of the Basque Country, UPV/EHU, and the collaboration of G. Ros in the development of the study.

## BIOGRAPHIES

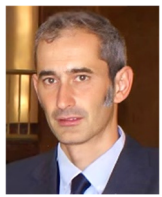

Name: **Javier**
Surname: **Bilbao**
Birthday: 1967
Birth Place: Spain
Master: Industrial Engineering from University of the Basque Country, Spain, in 1991
Doctorate: Ph.D. in Applied Mathematics, University of the Basque Country, Spain
The Last Scientific Position: Prof., Applied Mathematics Department, Engineering School of Bilbao, University of the Basque Country, Spain
Research Interests: Distribution overhead electrical lines compensation, Optimization of series capacitor batteries in electrical lines, Modelization of a leakage flux transformer, Losses in the electric distribution Networks, Artificial Neural Networks, Modelization of fishing trawls, E-learning, Noise of electrical wind turbines, Light pollution, Machine Learning, Computational Thinking
Scientific Publications: 76 Papers, 30 Books, 1 Patent, 48 Projects, 161 conference papers
Scientific Memberships: Bebras Community, IOTPE Organization

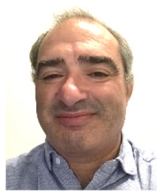

Name: **Eugenio**
Surname: **Bravo**
Birthday: 1967
Birth Place: Spain
Master: Industrial Engineering from University of the Basque Country, Spain, 1991
Doctorate: Ph.D. in Electrical Engineering, University of the Basque Country, Spain
The Last Scientific Position: Prof., Applied Mathematics Department, Engineering School of Bilbao, University of the Basque Country, Spain
Research Interests: distribution overhead electrical lines compensation, optimization of series capacitor batteries in electrical lines, modelization of a leakage flux transformer, losses in the electric distribution networks, artificial neural networks, modelization of fishing trawls, e-learning, noise of electrical wind turbines, computational thinking
Scientific Publications: 56 Papers, 25 Books, 1 Patent, 38 Projects, 135 conference papers
Scientific Memberships: Bebras Community, IOTPE Organization

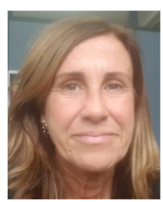

Name: **Olatz**
Surname: **Garcia**
Birthday: 1968
Birth Place: Spain
Master: Mathematics, University of the Basque Country, Spain
Doctorate: Science, Mathematics section, University of the Basque Country, Spain
The Last Scientific Position: Prof., Applied Mathematics Department, Engineering School of Bilbao, University of the Basque Country, Spain
Research Interests: e-learning, optimization of series capacitor batteries in electrical lines, Noise of electrical wind turbines, Machine Learning, Computational Thinking
Scientific Publications: 45 Papers, 25 Books, 1 Patent, 33 Projects, 111 conference papers
Scientific Memberships: Bebras Community, IOTPE Organization

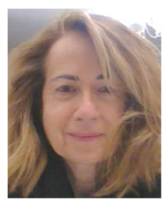

Name: **Carolina**
Surname: **Rebollar**
Birthday: 1963
Birth Place: Spain
Master: Mathematics, University of the Basque Country, Spain
Doctorate: Science, Mathematics section, University of the Basque Country, Spain
The Last Scientific Position: Prof., Applied Mathematics Department, Engineering School of Bilbao, University of the Basque Country, Spain
Research Interests: e-learning, Noise of electrical wind turbines, Machine Learning, Computational Thinking
Scientific Publications: 36 Papers, 24 Books, 1 Patent, 25 Projects, 87 conference papers
Scientific Memberships: Bebras Community, IOTPE Organization, SEMA